\documentclass[pre,aps,10pt,superscriptaddress,twocolumn,floatfix]{revtex4-2}

\usepackage[nice]{nicefrac}
\usepackage{graphicx,color}
\usepackage{amsmath,amssymb,bm,bbm}
\usepackage{amsmath}
\usepackage{braket}
\usepackage{bbold}

\usepackage{physics}
\usepackage{accsupp}

\usepackage[plainpages=false,pdfpagelabels,colorlinks=true,linkcolor=red,urlcolor=blue,citecolor=blue,pdftitle={},pdfauthor={},pdfdisplaydoctitle=true,pdfduplex=DuplexFlipLongEdge]{hyperref}

\definecolor{darkred}{rgb}{0.90,0.2,0.2}
\definecolor{darkgreen}{rgb}{0,0.60,.2}
\definecolor{darkblue}{rgb}{0.1,0.3,1}
\definecolor{grey}{cmyk}{0,0,0,0.25}
\definecolor{orange}{cmyk}{0,0.6,0.8,0}

\usepackage[T1]{fontenc}

\begin{document}
\title{Information Compression at Criticality}

\author{Simon Jiricek}
\affiliation{Department of Theoretical Physics, J. Stefan Institute, SI-1000 Ljubljana, Slovenia}
\affiliation{Department of Physics, Faculty of Mathematics and Physics, University of Ljubljana, SI-1000 Ljubljana, Slovenia\looseness=-1}

\author{Miroslav Hopjan}
\affiliation{Institute of Theoretical Physics, Wrocław University of Science and Technology, 50-370 Wrocław, Poland}

\author{Boris Altshuler}
\affiliation{Physics Department, Columbia University, 538 West 120th Street, New York, NY 10027}

\author{Vladimir Kravtsov}
\affiliation{International Centre for Theoretical Physics (ICTP), Strada Costiera 11, 34151 Trieste, Italy}

\author{Lev Vidmar}
\affiliation{Department of Theoretical Physics, J. Stefan Institute, SI-1000 Ljubljana, Slovenia}
\affiliation{Department of Physics, Faculty of Mathematics and Physics, University of Ljubljana, SI-1000 Ljubljana, Slovenia\looseness=-1}

\begin{abstract}
Highly excited quantum states at the critical boundary of ergodicity are known to deviate from thermal behavior, yet their dynamical properties remain poorly understood. Here, we uncover the complexity of quantum dynamics at criticality through the lens of intrinsic information compression in energy space. We show that the Hamiltonian spectrum can be systematically truncated, yielding a simplified description of the dynamics while preserving its essential features. Specifically, for both interacting and noninteracting systems, we demonstrate that a vanishing fraction of Hamiltonian eigenlevels suffices to reproduce the power-law decay of the survival probability. The resulting truncated spectrum exhibits a fractal structure characterized by a level-spacing distribution with a power-law tail, while its spectral form factor displays the same asymptotic power-law decay as the survival probability.
\end{abstract}

\maketitle

\textit{Introduction.---}%
The central goal of statistical physics is to reduce complexity by identifying the minimal ingredients required to describe universal phenomena.
In quantum many-body physics, for instance, tensor-network representations such as matrix product states~\cite{white92, Schollwoeck11} provide an efficient compression of weakly entangled many-body wavefunctions~\cite{Hastings_2007, Eisert_2010, Schollwoeck11}. 
Efficient descriptions are also available for other important classes of quantum states, including stabilizer states~\cite{gottesman1997, gottesman_98, scott_2004, chitambar_19} and Gaussian states~\cite{weedbrook_2012}.

Quantum dynamics in isolated many-body systems generally generate substantial entanglement, limiting the applicability of conventional wavefunction compression schemes. 
Yet, even at long times, the resulting states need not be maximally entangled. 
This is the case at the boundary of ergodicity, our focus here, where critical wavefunctions exhibit submaximal entanglement and a multifractal structure. 
Prominent examples include the Anderson localization transition~\cite{evers_mirlin_08, Rodriguez09, Rodriguez10} and interacting systems undergoing many-body ergodicity-breaking transitions~\cite{suntajs_vidmar_22, suntajs_deroeck_24, pawlik_zakrzewski_2024, swietek_2026}. 
These observations naturally raise the question of whether critical states admit efficient information-compression schemes. 
Certain progress in this direction has been made for structured random-matrix models, where toy wave functions have been constructed to reproduce the desired multifractal properties~\cite{detomasi_20, mirlin_fyodorov_96, Kravtsov2015}.

In this Letter, we introduce a new paradigm for information compression that operates in energy rather than Hilbert space. 
Instead of sampling many-body wavefunctions, we identify and sample the subset of Hamiltonian eigenlevels that controls the dynamics, leading to an emergent compression of the energy spectrum.

\begin{figure}[!t]
\centering
\includegraphics[width=\columnwidth]{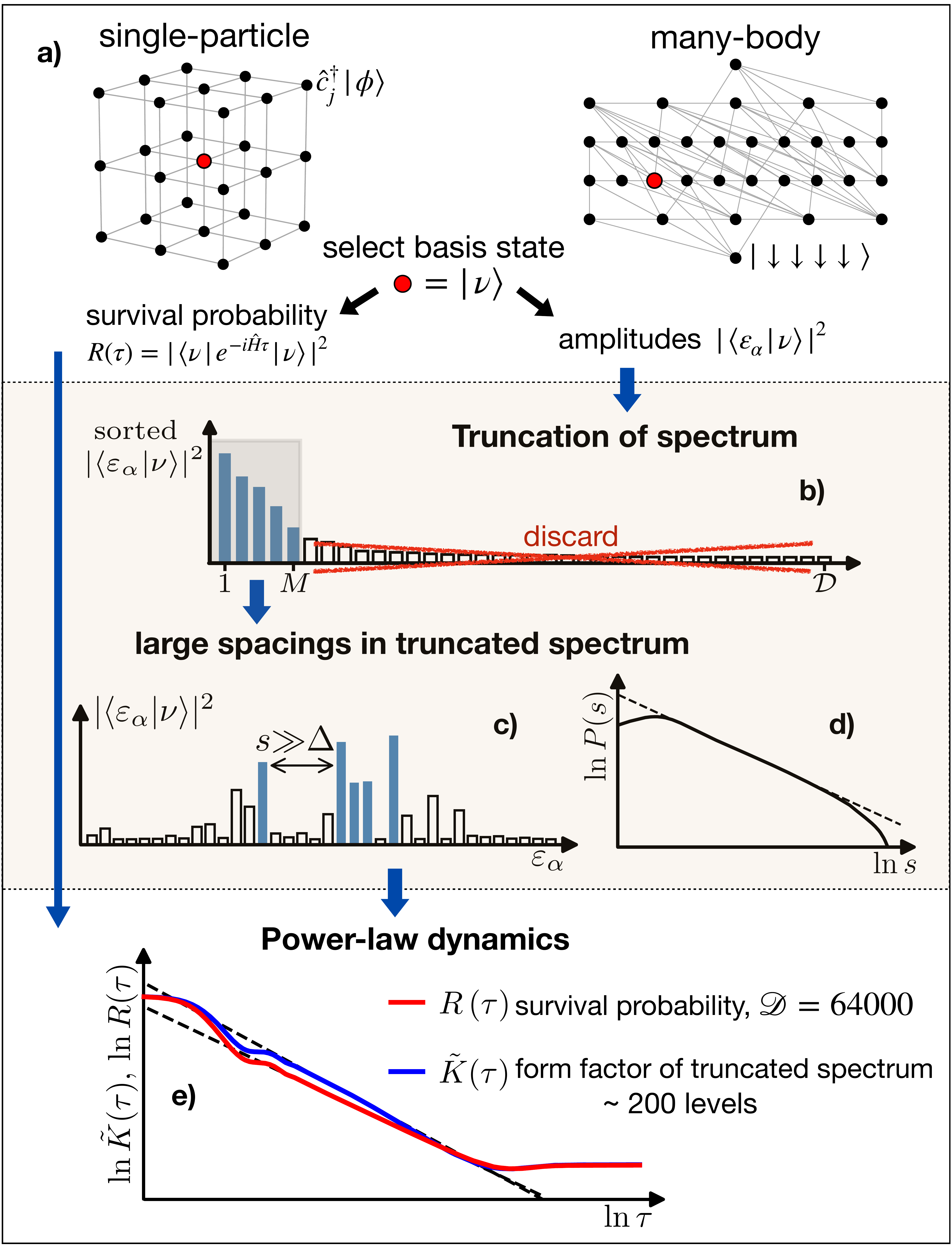}
\caption{Sketch of the truncation procedure. 
(a) We select a basis state $|\nu\rangle$, corresponding to a single lattice site for the single-particle models or a specific qubit configuration for the many-body models.
(b) We truncate the spectrum based on the amplitudes of the selected basis state by keeping only the $M$ largest contributions to the wavefunction, see the main text for details. 
In (c) we sketch how large spacings appear in the truncated spectrum giving rise to (d) a power-law distribution of spacings. 
(e) The truncated SFF $\tilde{K}(\tau)$, see Eq.~\eqref{eq:def_sff_trunc_spec}, shows that the information about a power-law decay of $R(\tau)$ is contained, for each basis state, in a vanishing fraction of energy levels.
}
\label{fig1}
\end{figure}

Studying both interacting and noninteracting models with either sparse or dense Hamiltonians, we show that an emergent compression of the Hamiltonian spectrum underlies the well-known power-law decay of the survival probability at criticality~\cite{Peres84, Chalker88, Chalker_90, Ketzmerick_92, Schofield_95, Ng_06, torresherrera_santos_15, tavora_torresherrera_16, bera_detomasi_18, hopjan_vidmar_23, Hopjan23b}. 
Remarkably, the fractal dimension of the truncated spectrum that faithfully reproduces the dynamics coincides with the fractal dimension of the eigenstates. 
Moreover, the truncated spectrum develops large gaps, leading to a level-spacing distribution with a power-law tail, while its spectral form factor exhibits the same power-law decay as the survival probability; see also Fig.~\ref{fig1}.

Our results have important implications for quantum dynamics at the transition between ergodic and nonergodic phases. In particular, they demonstrate that the critical dynamics are governed by only a vanishing fraction of the Hamiltonian eigenlevels.

\textit{Survival probability and SFF at criticality.---}%
Important perspective on the dynamics in critical quantum systems is given by the survival probability (also called return probability). The survival probability $R(\tau)$ for a given initial state $\ket{\nu}$ is given by the overlap with its time-evolved counterpart as
$R(\tau) = |\bra{\nu} e^{-i\hat{H}\tau}\ket{\nu}|^2$.
At criticality, $R(\tau)$ of typical basis states assumes a scale-invariant power-law in the mid-time regime, $R(\tau)\!\sim\! \tau^{-d_2}$~\cite{Peres84,Chalker88, Chalker_90, Ketzmerick_92, Schofield_95, Ng_06, hopjan_vidmar_23, Hopjan23b}, where the exponent of the decay is given by the wavefunction fractal dimension $d_2$~\cite{Chalker88, Chalker_90}.
In the Hamiltonian eigenbasis $\ket{\varepsilon_\alpha}$, $R(\tau)$ can be expressed as 
\begin{align}
\label{eq:survivial_in_H_basis}
    R(\tau) = \sum_{\alpha,\beta=1}^{\cal D} |\braket{\varepsilon_\alpha}{\nu}|^2 \,|\braket{\varepsilon_\beta}{\nu}|^2\, e^{-i(\varepsilon_\alpha - \varepsilon_\beta) \tau}\;,
\end{align}
where $\mathcal{D}$ is the corresponding Hilbert-space dimension, e.g., ${\cal D}=L^d$ for a single particle on a hypercube with linear size $L$ in dimension $d$, or ${\cal D} = 2^L$ for a system of $L$ qubits.
$R(\tau)$ in Eq.~\eqref{eq:survivial_in_H_basis} carries certain similarities with the spectral form factor (SFF)~\cite{berry_85, sieber_richter_01, mueller_heusler_04, kos_ljubotina_18, bertini_kos_18, Chan18, chan_deluca_18a, suntajs_bonca_20a, sierant_delande_20, Prakash21, suntajs_prosen_21, suntajs_vidmar_22, roy_prosen_20, roy_mishra_22, Joshi22, matsoukasRoubeas_beau_23, dag_mistakidis_23, fritzsch_prosen_24, dong_zhang_25}, which measures correlations in the Hamiltonian spectrum,
\begin{align}
\label{eq:def_sff}
    K(\tau) = \frac{1}{\mathcal{D}^2}\, \Bigl|\sum_{\alpha=1}^{\mathcal{D}} \, e^{-i\,\varepsilon_\alpha \tau} \Bigr|^2\,,
\end{align}
with $\mathcal{D}^2$ acting as a normalization factor.

Despite the similarities between $R(\tau)$ and $K(\tau)$, these two quantities exhibit distinct dynamics at criticality: while $R(\tau)$ exhibits a power-law decay~\cite{Peres84,Chalker88, Chalker_90, Ketzmerick_92, Schofield_95, Ng_06, hopjan_vidmar_23, Hopjan23b}, $K(\tau)$ exhibits a broad plateau~\cite{suntajs_prosen_21, suntajs_prosen_23, suntajs_vidmar_22, Hopjan23b, Jiricek_26} whose value is given by the spectral compressibility~\cite{Chalker96a, Jiricek_26}. Possible common features in the dynamics of $R(\tau)$ and $K(\tau)$ are therefore not immediately obvious. 
Recently, the idea was put forward~\cite{Altshuler23}, that the emergence of the power-law in $R(\tau)$ is governed by contributions of only a small set of eigenlevels with considerable weight $|\braket{\varepsilon_\alpha}{\nu}|^2$, while the correlations between different wavefunction projections are negligible. Subsequently, the authors conjectured a connection between the survival probability and the spectral form factor over the reduced set of eigenvalues.
Building on this ideas, we introduce below a simple and powerful truncation procedure that singles out the eigenlevels that govern quantum dynamics.
This opens doors towards establishing the emergent spectral compression at criticality, see also Fig.~\ref{fig1}, and further allows us to demonstrate the connection between the survival probability and the SFF.

\begin{figure}[!t]
\centering
\includegraphics[width=\columnwidth]{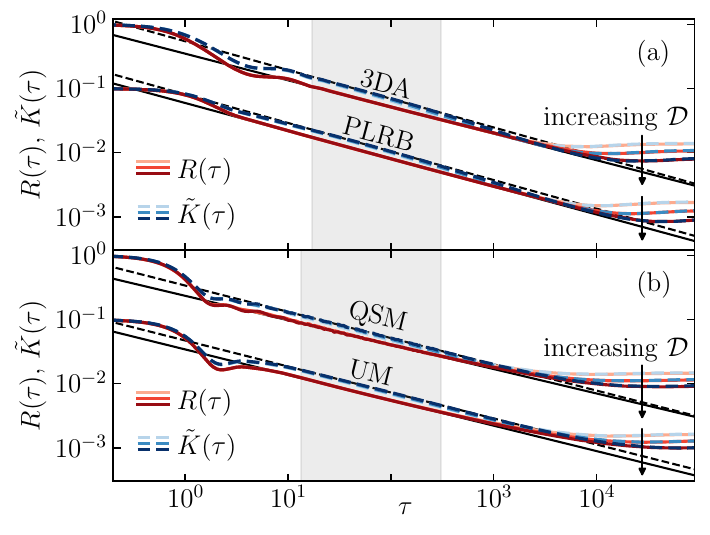}
\caption{Comparison of the survival probability $R(\tau)$ [red solid lines] and the truncated SFF $\tilde{K}(\tau)$ [blue dashed lines].
(a) The 3d Anderson Model (3DA) for $\mathcal{D}=26^3, 32^3, 40^3$ and the power-law random banded model (PLRB) for $\mathcal{D}=16\text{k}, 32\text{k}, 64\text{k}$, (b) the quantum sun model (QSM) and the ultrametric model (UM), both for $\mathcal{D}=2^{14}, 2^{15}, 2^{16}$. 
Increasing Hilbert-space dimension is indicated by the arrows and deeper colors. The data for different models is shifted by a factor of 10.
We carry out power-law fits in the shaded region to the functions $R(\tau)\propto \tau^{-d_2'}$ [black solid lines] and $\tilde{K}(\tau) \propto \tau ^{-\tilde{d}_2}$ [black dashed lines]. 
For the 3DA, PLRB, QSM and UM, respectively, we obtain $d_2' = 0.42, 0.43, 0.38, 0.40$ and $\tilde{d}_2 = 0.45, 0.45, 0.41, 0.41$.
}
\label{fig2}
\end{figure}

\textit{The truncation procedure.---}%
The main idea of the truncation is to sample $M$ eigenvalues $\tilde{\varepsilon}_\alpha$, with $M \ll {\cal D}$, which govern the dynamics of the survival probability $R(\tau)$, while neglecting the correlations of the wavefunction amplitudes in $R(\tau)$.
This set of eigenvalues, denoted by $\mathcal{E} = \{\tilde{\varepsilon}_\alpha\}$, corresponds to eigenstates that have the largest overlap, $|\braket{\varepsilon_\alpha}{\nu}|^2$, with the initial state $\ket{\nu}$, see Fig.~\ref{fig1}(b). In other words, the eigenvalues $\tilde{\varepsilon}_\alpha$ are the most probable samples when measuring the total energy $\hat{H}$ of state $\ket{\nu}$.
In this way, one approximates $R(\tau)$ in Eq.~\eqref{eq:survivial_in_H_basis} as
$\tilde R(\tau) = \sum_{\tilde{\varepsilon}_\alpha, \tilde{\varepsilon}_\beta\in \mathcal{E}} w^2\, e^{-i(\tilde{\varepsilon}_{\alpha} \,- \,\tilde{\varepsilon}_\beta) \tau}$,
where the wavefunction amplitudes are replaced by a constant weight $w$. 
As a minimal condition for $\tilde R(\tau)$ to be a reasonable approximation of $R(\tau)$, it should (i) equal the initial state value $R(\tau\!\!=\!\!0) =1$, giving rise to the condition $w=1/M$, and (ii) equal to the infinite time value $R(\tau \!\xrightarrow{} \!\infty) = I_2$, where $I_2$ is the inverse participation ratio, $I_2=\sum_\alpha |\braket{\varepsilon_\alpha}{\nu}|^4$, of the state $\ket{\nu}$ in the eigenbasis of the Hamiltonian. 
This gives rise to the condition $Mw^2\approx I_2$, i.e., $M \approx 1/I_2$. 
In the actual truncation procedure, we set $M = \mathrm{round}(1/I_2)$,
i.e., $M$ is the closest integer to $I_2^{-1}$.
One then realizes that the approximate survival probability $\tilde{R}(\tau)$ is nothing else but the SFF over the truncated spectrum $\tilde{K}(\tau)$ (shortly, truncated SFF), provided the initial value is normalized to unity,
\begin{align}
\label{eq:def_sff_trunc_spec}
    \tilde{K}(\tau) = \frac{1}{M^2}\,\Bigl|\sum_{\tilde{\varepsilon}_{\alpha} \in \mathcal{E}} e^{-i \tilde{\varepsilon}_{\alpha} \tau } \Bigr|^2\,.
\end{align}
As suggested by Fig.~\ref{fig1}(e), the truncated SFF $\tilde{K}(\tau)$ from Eq.~\eqref{eq:def_sff_trunc_spec} is indeed a very accurate approximation for the exact survival probability $R(\tau)$ from Eq.~\eqref{eq:survivial_in_H_basis}.
A detailed numerical analysis will follow in Fig.~\ref{fig2}.

We note that $M\propto I_2^{-1}$ implies the scaling $M \propto {\cal D}^{\,d_2}$, where $d_2$ is the wavefunction fractal dimension obtained from the scaling of the IPR, $I_2\propto {\cal D}^{\,-d_2}$.
Hence, if $d_2<1$, as found at the ergodicity breaking transition of all the models studied here, we have $M/{\cal D}\!\to\! 0$, i.e., the truncated spectrum consists of a measure zero of all eigenvalues.

\textit{Models.---}%
We consider four paradigmatic models hosting an ergodicity breaking transition.
Two of them are quadratic, expressed as 
$\hat H = \sum_{i,j=1}^{\cal D} h_{ij} \hat c_i^\dagger \hat c_j$, where ${\cal D}$ is the single-particle Hilbert space and $\hat{c}^\dagger_i\,$($\hat{c}_i$) are creation (annihilation) operators for spinless fermions on lattice sites $i$.
The first quadratic model is the three-dimensional Anderson model (3DA) with ${\cal D}=L^3$~\cite{Anderson_58}, for which the only non-zero matrix elements are $h_{\langle i,j \rangle}=-t$ denoting the nearest neighbor hoppings on the cubic lattice (we set $t=1$), and $h_{ii}$ denoting the random disorder potential independently drawn from a box distribution, $h_{ii}\in[-W/2,W/2]$.
The second quadratic model is the power-law random banded model (PLRB)~\cite{mirlin_fyodorov_96, evers_mirlin_08}, for which the Hamiltonian matrix is dense, i.e.,
$h_{ij} = \mu_{ij}/[1+(\frac{\mathcal{D}}{\pi}\sin(|i-j|\frac{\pi}{\mathcal{D}})/b)^{2a}]^{1/2}$,
and $\mu_{ij}$ are elements of a matrix $\mathbf{\mu}$ drawn from the Gaussian orthogonal ensemble.
We study both models at the single-particle ergodicity breaking transition, at which the Hamiltonian eigenstates in the middle of the spectrum are fractal, i.e., $d_2 < 1$.
In the 3DA, this corresponds to the well-known localization transition that emerges at $W_\text{c} = 16.5$~\cite{Slevin_2014, suntajs_prosen_21}, while in the PLRB, the transition emerges at $a_c=1$~\cite{mirlin_fyodorov_96, Hopjan23b}.
We set $b=0.42$ in the PLRB such that $d_2$ in both models are nearly identical.

The two other models are defined in a many-body Hilbert space of $L$ qubits, i.e., ${\cal D} = 2^L$.
The first of the two is the quantum sun model (QSM)~\cite{deroeck_huveneers_17, suntajs_vidmar_22, suntajs_deroeck_24}, an interacting model consisting of an all-to-all interacting random dot of fixed size $N=3$, and $L'=L-N$ qubits outside the dot that couple to the dot in a hierarchical order. 
The Hamiltonian reads
\begin{align}
\label{eq:def_qsun}
    \hat{H}_\mathrm{QSM} = \hat{H}_\mathrm{dot}  + g_0 \sum_{j=0}^{L'-1} \alpha^{u_j} \hat{S}_{n(j)}^x \hat{S}_j^x + \sum_{j=0}^{L'-1} h_j \hat{S}_j^z, 
\end{align}
where the first term describes the interactions inside the quantum dot, the second term contains the coupling from dot to outside qubits, and the last term adds random disorder fields in $z$-direction for each outside qubit. 
The second model we consider in a many-body Hilbert space is the ultrametric model (UM)~\cite{fyodorov_ossipov_09, rushkin_ossipov_11, bogomolny_giraud_11, bogomolny_sieber_18, vansoosten_warzel_18, suntajs_deroeck_24}, described by the Hamiltonian 
\begin{align}
\label{eq:def_um}
    \hat{H}_{\rm UM} = \hat{H}_0 + J \sum_{k=1}^L \alpha^k \hat{H}_k \,,
\end{align}
where each of the $\hat{H}_k$ describes blocks of random matrices, whose size decreases with $k$, while their magnitude is controlled by $\alpha$. 
In End Matter, we provide further details about the models in Eqs.~\eqref{eq:def_qsun} and~\eqref{eq:def_um}, as well as about the numerical averages over different basis states in all models under investigation.

While the underlying Hamiltonian matrix is sparse for the QSM, it is dense for the UM~\cite{suntajs_deroeck_24}. The many-body interpretation of the latter is therefore a choice of interpretation, motivated by recent studies that showed how both the QSM and the UM exhibit many similar properties when expressed in the same Hilbert space~\cite{suntajs_deroeck_24, Hopjan23b, swietek_2026}.
Here we study both models at the ergodicity breaking transition, which emerges at $\alpha_c=1/\sqrt{2}$ and gives rise to multifractal eigenstates in the qubit configuration basis~\cite{suntajs_deroeck_24}.
In the QSM, however, we slightly modify $\alpha_c$ to $\alpha_c=0.734$~\cite{swietek_hopjan_25}, which is a transition point observed for a given set of model parameters in finite systems. 

\textit{Numerical test of the truncation procedure.---}%
In Fig.~\ref{fig2} we test the relevance of the truncation procedure by comparing the survival probability $R(\tau)$ to the truncated SFF $\tilde{K}(\tau)$.
Remarkably, in all four models under investigation, the dynamics at the critical point are governed by a power-law decay, and the agreement between $R(\tau)$ and $\tilde{K}(\tau)$ is nearly perfect.

We parametrize in Fig.~\ref{fig2} the dynamics of both quantities as power-laws $R(\tau)\propto \tau^{-d_2'}$ [black solid lines] and $\tilde{K}(\tau) \propto \tau ^{-\tilde{d}_2}$ [black dashed lines].
We observe that the exponents $d_2'$ are very close to the exponents $\tilde{d}_2$, see the caption of Fig.~\ref{fig2}.
We argue that small differences between the exponents stem from small deviations between $R(\tau)$ and $\tilde{K}(\tau)$ at short times.
However, since the long-time values of $R(\tau)$ and $\tilde{K}(\tau)$ are equal by definition, we then also expect $\tilde{d}_2\to d_2'$ with increasing the system size $\cal D$.
We note that the exponents $d_2'$ and $\tilde{d}_2$ also agree to high accuracy with the wavefunction fractal dimensions $d_2$, as anticipated from~\cite{hopjan_vidmar_23, Hopjan23b}, i.e., $d_2\approx d_2'\approx \tilde{d}_2$.

The main message from Fig.~\ref{fig2} is that the agreement between the survival probability and the truncated SFF is nearly perfect, despite the ratio $M/{\cal D}$ of the retained levels after truncation vanishes in the thermodynamic limit.
Specifically, $M/{\cal D}< 10^{-2}$ for the largest system sizes considered in Fig.~\ref{fig2}.

\begin{figure}[!t]
\centering
\includegraphics[width=\columnwidth]{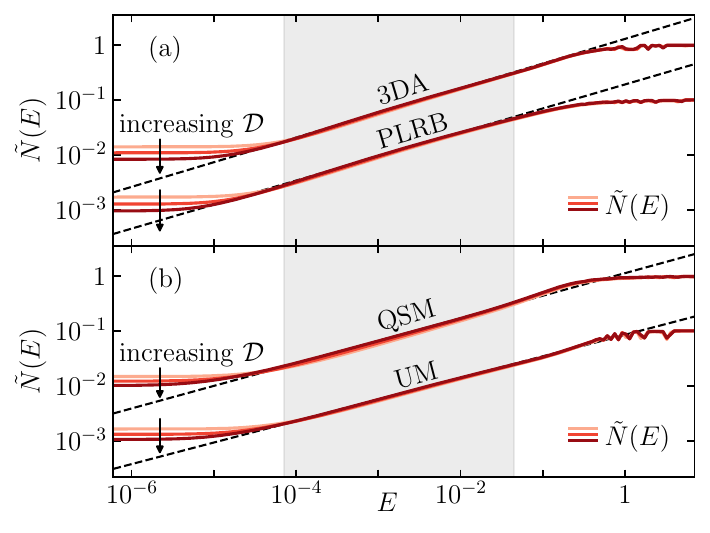}
\caption{
Box counting method in the truncated spectrum.
(a) The 3d Anderson Model (3DA) for $\mathcal{D}=26^3, 32^3, 40^3$ and the power-law random banded model (PLRB) for $\mathcal{D}=16\text{k}, 32\text{k}, 64\text{k}$, (b) the quantum sun model (QSM) and the ultrametric model (UM), both for $\mathcal{D}=2^{14}, 2^{15}, 2^{16}$. 
Increasing Hilbert-space dimension is indicated by the arrows and deeper colors. The data for different models is shifted by a factor of 10.
We carry out power-law fits in the shaded region to the function $\tilde{N}(E)\propto E^{\,\tilde{\delta}_2}$ [black dashed lines].
For the 3DA, PLRB, QSM and UM, respectively, we obtain the exponents $\tilde{\delta}_2=0.45, 0.44, 0.41, 0.39$. 
}
\label{fig3}
\end{figure}

\textit{Fractality of the truncated spectrum.---}%
Finally, we study statistical properties of the truncated spectrum.
We first calculate the spectral fractal dimension using the box counting method. 
In the latter, we define the scaling function of the truncated spectrum,
\begin{equation}
\label{eq:def_box_count}
     \tilde{N}_{}^{}(E)  = \frac{1}{M^{2}}\sum_{B\neq B_\varnothing} \Bigl(\sum_{\tilde{\varepsilon}_\alpha \in B} 1 \Bigr)^{2} \;,
\end{equation}
where the first sum runs over non-empty boxes, $B\neq B_\varnothing$, and the second sum counts the number of levels in each box of width $E$, see End Matter for details.
The exponent $\tilde{\delta}_2$ of the power-law behavior, $\tilde{N}(E)\propto E^{\,\tilde{\delta}_2}$, can be interpreted as the spectral fractal dimension~\cite{hopjan25}.
In Fig.~\ref{fig3} we observe $\tilde{\delta}_2<1$, suggesting that the truncated spectrum is fractal. We numerically tested (not shown) that similar spectral fractal dimensions are obtained for exponents in Eq.~\eqref{eq:def_box_count} other than 2, suggesting that the truncated spectrum may be a mono-fractal. 

It is not immediately obvious whether and how the spectral fractal dimension $\tilde{\delta}_2$ is related to the wavefunction fractal dimension $d_2$.
It was argued~\cite{Ketzmerick_92, Altshuler23} that $\tilde{\delta}_2$ is related to the power-law exponent of the corresponding SFF, i.e., $\tilde{\delta}_2 \approx \tilde{d}_2$~\footnote{It was shown in Ref.~\cite{Ketzmerick_92} that ${\tilde{\delta}_2}$ is connected to the slope of the integrated SFF, i.e., $\tilde{C}(\tau)=\frac{1}{\tau}{\protect\int}_{0}^{\tau}\tilde{K}(\tau')d\tau'\propto \tau^{-\tilde{\delta}_2}$. Since $\tilde{\delta}_2<1$, the slopes of $\tilde{K}(\tau)$ and $\tilde{C}(\tau)$ are proportional, i.e., $\tilde{K}(\tau) \propto \tilde{C}(\tau)$, and thus $\tilde{\delta}_2 \approx \tilde{d}_2$.}.
Based on our results for the similarity between the truncated SFF and the survival probability, see Figs.~\ref{fig2} and~\ref{fig3} and the numbers reported in the figure captions, one may hence conjecture that $\tilde{\delta}_2$ is also related to the wavefunction fractal dimension $d_2$, since $ \tilde{\delta}_2 \approx \tilde{d}_2 \approx d_2'\approx d_2$.

\begin{figure}[!t]
\centering
\includegraphics[width=\columnwidth]{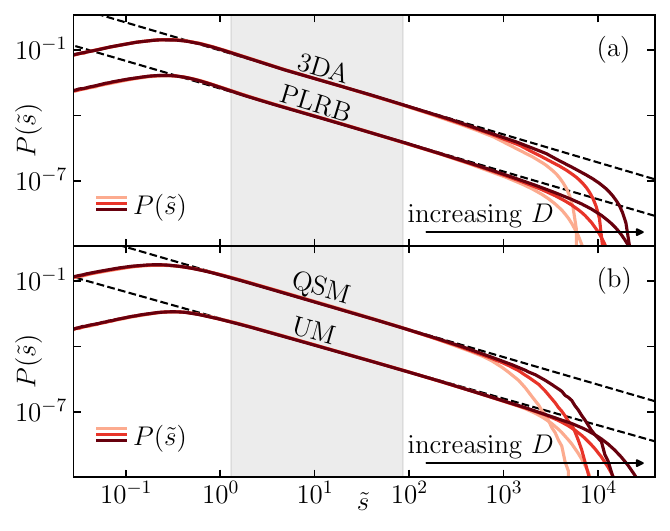}
\caption{
Distributions $P$ of level spacings $\tilde{s}$ of the truncated spectrum.
(a) The 3d Anderson Model (3DA) for $\mathcal{D}=26^3, 32^3, 40^3$ and the power-law random banded model (PLRB) for $\mathcal{D}=16\text{k}, 32\text{k}, 64\text{k}$, (b) the quantum sun model (QSM) and the ultrametric model (UM), both for $\mathcal{D}=2^{14}, 2^{15}, 2^{16}$. 
Increasing Hilbert-space dimension is indicated by the arrows and deeper colors. The data for different models is shifted by a factor of 100.
We carry out power-law fits in the shaded region to the function $P(\tilde{s}) \propto \tilde{s}\,^{-(1+\tilde{\Delta}_2)}$ [black dashed lines].
For the 3dA, PLRB, QSM and UM, respectively, we obtain the exponents $\tilde{\Delta}_2 = 0.28, 0.27, 0.26, 0.22$.
}
\label{fig4}
\end{figure}

Another manifestation of fractality of the truncated spectrum can be observed in the level spacing distribution $P(\tilde{s})$ of the nearest level gaps $\tilde{s}_\alpha=\tilde{\varepsilon}_\alpha - \tilde{\varepsilon}_{\alpha-1}$. The distribution exhibits a power-law decay that we parametrize as $P(\tilde{s}) \propto \tilde{s}^{-(1+\tilde{\Delta}_2)}$, with the exponent $\tilde{\Delta}_2<1$, see Fig.~\ref{fig4}. The question naturally emerges, whether the exponent $\tilde{\Delta}_2$ is related or equal to the exponents $d_2$ and $\tilde{\delta}_2$.
Our results for $\tilde{\delta}_2$ and $\tilde{\Delta}_2$, see the captions of Figs.~\ref{fig3} and~\ref{fig4}, suggests that the exponents may not agree, at least for the system sizes under investigation.

So far, only a few studies attempted to establish theoretical predictions for the relationship between the power-law exponents $\tilde{d}_2$ and $\tilde{\Delta}_2$ at criticality. 
Recently, it was argued~\cite{Altshuler23} that the power-law exponent $\tilde{\Delta}_2^\text{CS}$ of the level spacing distribution of a Cantor set (CS), $P_\text{CS}(s) \propto s^{-(1+\tilde{\Delta}_2^\text{CS})}$, matches the power-law exponent $\tilde{d}_2^\text{CS}$ of the corresponding SFF $\tilde{K}_\text{CS}(\tau) \propto \tau ^{-\tilde{d}_2^\text{CS}}$, i.e., $\tilde{\Delta}_2^\text{CS} = \tilde{d}_2^\text{CS}$. It was further shown that the same relations hold for random fractal sets \cite{Altshuler23} consisting of uncorrelated levels.  
In our case, we argue that the absence of such quantitative agreement, $\tilde{\Delta}_2^\text{} \neq \tilde{d}_2^\text{} \approx \tilde{\delta}_2^\text{} $, 
is caused by strong correlations between level spacings in the truncated spectrum, see End Matter.
There exist, however, physical models, such as the quasiperiodic 1D Aubry-André model at the localization transition \cite{Aubry80}, in which, at the level of the global spectrum (i.e., without truncation), the exponent from the level spacing distribution is quantitatively close to the spectral fractal dimensions \cite{Ketzmerick_92, Geisel_91, hopjan_vidmar_23}, see End Matter for details. Yet we note that in the latter case we deal with the built-in fractality of the spectrum, whereas in our present case this fractality is an emergent property that is revealed by truncation.

\textit{Conclusions.---}%
In this Letter, we showed that the critical dynamics of both interacting and noninteracting systems are encoded in a vanishing fraction of Hamiltonian eigenlevels, defining a truncated spectrum. This spectrum is fractal and exhibits a spectral form factor with a power-law decay whose exponent closely matches the eigenstate fractal dimension. Furthermore, its level-spacing distribution develops a power-law tail, revealing a distinctive spectral signature of ergodicity-breaking transitions beyond the paradigms of Gaussian random-matrix ensembles and Poisson statistics. Our results establish intrinsic information compression in energy space as a fundamental property of quantum dynamics at criticality, complementing existing wavefunction compression schemes and opening a new avenue for understanding and describing complex quantum dynamics.

\acknowledgements
We acknowledge support from the Slovenian Research and Innovation Agency (ARIS), Research core funding Grants No.~P1-0044, N1-0273 and J1-50005, as well as the Consolidator Grant Boundary-101126364 of the European Research Council (ERC) (S.J, M.H. and L.V.).
M. H. acknowledges support from the Polish National Agency for Academic Exchange (NAWA)’s Ulam Programme (project BNI/ULM/2024/1/00124).
We gratefully acknowledge the High Performance Computing Research Infrastructure Eastern Region (HCP RIVR) consortium~\cite{vega1} and European High Performance Computing Joint Undertaking (EuroHPC JU)~\cite{vega2}  for funding this research by providing computing resources of the HPC system Vega at the Institute of Information sciences~\cite{vega3}.

\bibliographystyle{biblev1}
\bibliography{references, references1, references2}

\onecolumngrid
\begin{center}
{\large \bf End Matter}\\
\end{center}
\twocolumngrid

{\it Appendix A: Details of the many-body models.---}%
In the quantum sun model Hamiltonian, Eq.~\eqref{eq:def_qsun}, the action of the random dot is given as $\hat{H}_\mathrm{dot} \equiv \frac{\gamma}{\sqrt{2^N+1}} \hat{R}_{\,\mathcal{H}_\mathrm{dot}} \otimes \mathbb{1}_{\,\mathcal{H}_\mathrm{out}}$, where $R$ is a $2^N \times 2^N$ matrix from the Gaussian orthogonal ensemble (GOE) and the prefactor is chosen such that $\gamma$ controls the bandwidth of the dot. The second term in Eq.~\eqref{eq:def_qsun} describes an interaction term in form of a hierarchical coupling of a randomly selected spin $n_{(j)}$ inside of the dot and the outside spin $j$ where the average strength of the random coupling is determined by $\alpha^j$ since $u_j=j + [-\zeta,\zeta]$ for all spins except the first where $u_0=0$. We fix the strength of random fluctuations in the coupling to $\zeta\! = 0.2$ throughout this work. The last term in Eq.~\eqref{eq:def_qsun} describes random fields $h_j$ acting on the spins $z$-components, with $h_j = W + [-\delta_W,  \delta_W] $, with $W\!=\!1$ and $\delta_W=0.5$. We are focusing on the parameters $g_0\!=\!1$ and $\gamma=1$ for which the critical point for numerically accessible system sizes was determined as $\alpha_c \approx 0.734$~\cite{swietek_hopjan_25}. At these parameters, the critical point exhibits fractal dimensions close to those of the 3d Anderson transition.

In the ultrametric model Hamiltonian, Eq.~\eqref{eq:def_um}, the $k$-th block is given as
$\hat{H}_k = \hat{R}/\sqrt{2^{N+k}+1}$,
where $R$ denotes a $2^{N+k} \times 2^{N+k} $ GOE matrix. For every $k$, there are $2^{L-k}$ independently drawn blocks, while the first block entry is placed on the diagonal of the full Hamiltonian~\cite{suntajs_deroeck_24}. The model exhibits an ergodicity breaking phase transition at $\alpha_c=1/\sqrt{2}$. The fractal dimensions at the transition depend on the additional parameter $J$ and are close to those in the 3d Anderson model for the value $J=0.27$ chosen in this work.

\begin{figure}[b]
\centering
\includegraphics[width=\columnwidth]{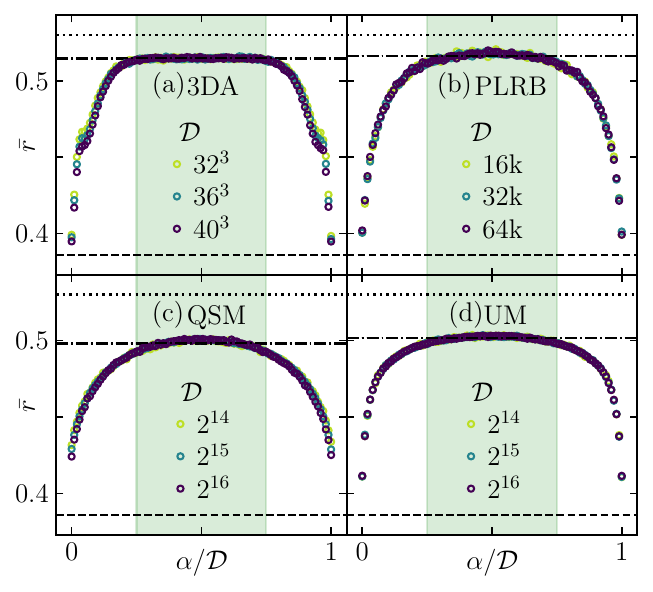}
\caption{Gap ratio in the global spectrum without truncation. The green shaded area indicates the 50\% of states we regard as critical states, the rest posing the mobility edge. The dotted lines indicate the GOE value $r_\text{GOE}\approx 0.53$, while the Poisson value $r_\text{Poisson} \approx 0.386$ is indicated by dashed lines. The dash-dotted lines correspond to the average gap ratio in the green shaded area. All results are averaged over different disorder realizations. Results are displayed for (a) the 3d Anderson Model, (b) the Power-law random banded model, (c) the Quantum sun model, (d) the Ultrametric model.}
\label{fig5}
\end{figure}

{\it Appendix B: Avoiding the mobility edge.---}%
All the models under consideration exhibit a mobility edge, i.e., at all parameter regimes below the critical point as well as at the critical point itself, localized states at the edges of the spectrum coexist with delocalized or critical states in the middle of the spectrum. Therefore, to unveil genuine properties of criticality in these systems, the contributions from the localized spectral edges must be suppressed.

To achieve this, we proceed in two steps. First, we analyze the precise location of the mobility edge in our systems by extracting gap ratios $r_i=\frac{\min (\varepsilon_{i+1} - \varepsilon_i, \varepsilon_i - \varepsilon_{i-1}) }{\max (\varepsilon_{i+1} - \varepsilon_i, \varepsilon_i - \varepsilon_{i-1})}$ across the energy spectra and plotting them as a running average, as displayed in Fig.~\ref{fig5}. The average gap ratio is a useful tool to distinguish localized states, with a value close to the gap ratio of Poisson statistics $r_\text{Poisson}\approx 0.386$, from delocalized states with a gap ratio close to the GOE value $r_\text{GOE}\approx 0.53$~\cite{Atas_13}, or critical states with intermediate values $r_\text{Poisson} < r< r_\text{GOE}$. In all four models we find that the 50\% of states around the middle of the spectrum yield similar intermediate values of the gap ratio, while outside of this range, the values drop down towards the Poisson limit. Therefore we approximately determine the window of critical states as the 50\% of states in the middle of the spectrum.

In the analysis throughout this work, we concern with basis states $\ket{\nu}$ and their projections to the energy eigenbasis $|\bra{\nu}  \ket{\varepsilon_\alpha}|^2$, which in principle involve all energy levels, also these at the localized edges of the spectrum. Therefore we impose an acceptance criterion for every basis state, in which a basis state is accepted only if the contributions of amplitudes $|\bra{\nu}  \ket{\varepsilon_\alpha}|^2$ from the edges of the spectra sum up to less than 40\% of the norm. This way we can avoid basis states $\ket{\nu}$ that lie close to a localization center of a localized edge state and have vanishing projections to any other eigenstates. We find that the acceptance criterion is passed by approximately 50\% of basis states, which we use for the analysis in this work.

\begin{figure}[!t]
\centering
\includegraphics[width=\columnwidth]{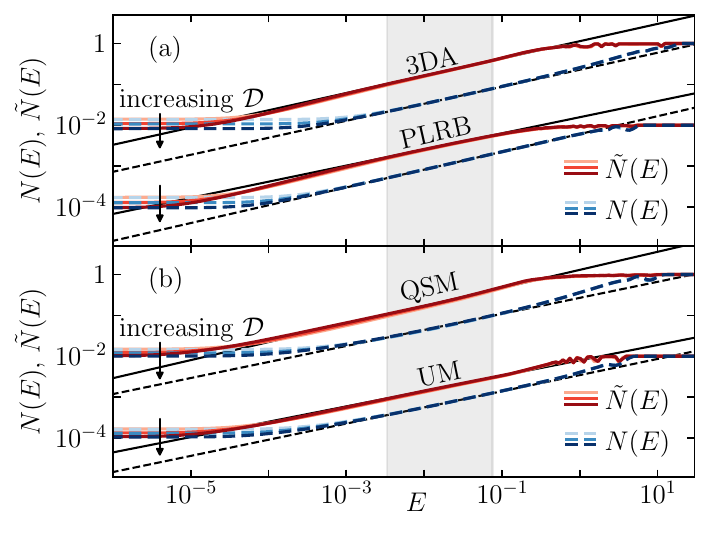}
\caption{
Comparison of box-counting methods for the full wavefunctions and the truncated spectrum, i.e., $N(E)$ and $\tilde{N}(E)$, respectively. 
(a) The 3d Anderson Model (3DA) for $\mathcal{D}=26^3, 32^3, 40^3$ and the power-law random banded model (PLRB) for $\mathcal{D}=16\text{k}, 32\text{k}, 64\text{k}$, (b) the quantum sun model (QSM) and the ultrametric model (UM), both for $\mathcal{D}=2^{14}, 2^{15}, 2^{16}$.
Increasing Hilbert-space dimension is indicated by the arrows and deeper colors. The data for different models is shifted by a factor of 100.
We carry out power-law fits in the shaded region to the functions $\tilde{N}(E)\propto E^{\,\tilde{\delta}_2}$ [black solid lines] and $N(E) \propto E ^{\,\delta_2}$ [black dashed lines]. 
For the 3DA, PLRB, QSM and UM, respectively, we obtain $\tilde{\delta}_2=0.45, 0.44, 0.41, 0.39$ and $\delta_2 = 0.42, 0.44, 0.39, 0.40$.
}
\label{fig6}
\end{figure}

{\it Appendix C: Box counting method.---}%
To define the box-counting method for the truncated spectrum, we start from the box-counting method for the complete set of wavefunction amplitudes and subsequently follow the same arguments as for the survival probability in the main text. The box-counting for a basis state $\ket{\nu}$ is defined as 
\begin{equation}
\label{eq:def_box_count_wavefunction}
N(E) = \sum_{B\neq B_\varnothing} \Bigl(\,\sum_{\varepsilon_\alpha \in B} |\braket{\varepsilon_\alpha}{\nu}|^2 \Bigr)^2,
\end{equation}
where the eigenvalues $\varepsilon_\alpha$ are counted in boxes of width $E$ and the first sum runs over all non-empty boxes. Averaged over different states and Hamiltonian realizations, a scaling $N(E)\propto E^{\,\delta_2}$ is observed, with an exponent $\delta_2$ that is close to the slope of survival probability $d_2'$~\cite{Ketzmerick_92}.

In the next step, we approximate Eq.~\eqref{eq:def_box_count_wavefunction} by replacing $|\braket{\varepsilon_\alpha}{\nu}|^2$ with a constant $w$ for the eigenvalues that remain after the truncation and zero otherwise. Imposing that the limiting values for $N(E \! \xrightarrow{} \! 0) = I_2$ and $N(E \! \xrightarrow{} \! \infty) = 1$ should remain unchanged, results in $w=1/M$, as we obtained for the emergence of the SFF over the truncated spectrum from the survival probability. Hence we arrive at Eq.~\eqref{eq:def_box_count} for the box-counting function $\tilde{N}(E)$ in the truncated spectrum.
To provide a further test of the truncation procedure, we provide in Fig.~\ref{fig6} a comparison between the wavefunction box-counting method and the box-counting in the truncated spectrum. Indeed we observe a similar power-law decay with almost identical exponents provided in the caption of Fig.~\ref{fig6}.

\begin{figure}[t!]
 \centering
 \includegraphics[width=\columnwidth]{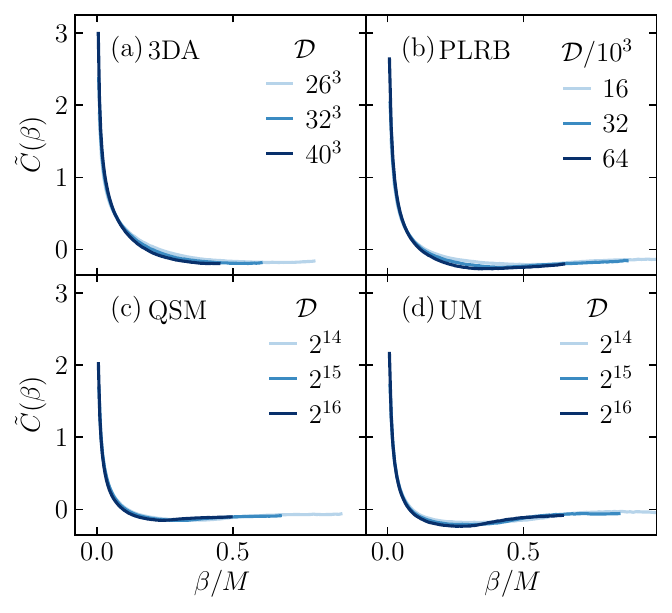}
 \caption{Correlations of spacings in the truncated spectrum, as defined in Eq.~\eqref{eq:def_corr} for the truncated spectrum. The $x$-axes is scaled by the average number of remaining levels in the truncated spectrum. Results are displayed for (a) the 3d Anderson Model, (b) the Power-law random banded model, (c) the Quantum sun model, (d) the Ultrametric model.}
 \label{fig7}
 \end{figure}

 {\it Appendix D: Remnant correlations in the truncated spectrum.---}%
 We define the correlation function for spacings $\tilde{s}$ between consecutive levels in the truncated spectrum as:
 \begin{align}
 \label{eq:def_corr}
     \tilde{C}(\beta) = \frac{\langle \tilde{s}_{\alpha + \beta}\, \tilde{s}_\alpha \rangle }{\langle \tilde{s}_{\alpha + \beta}\rangle \langle \tilde{s}_\alpha \rangle  } - 1.
 \end{align}
 In Fig.~\ref{fig7} we exemplarily show how rapidly decaying correlations in the spacings of the truncated spectrum remain present in all four different models.  

{\it Appendix E: Comparison to quasiperiodic models.---}%
An example of a  model whose global spectrum exhibits properties of an ideal fractal, i.e., $\tilde{\Delta}_2^\text{} \approx \tilde{d}_2^\text{}$, is the quasiperiodic 1d Aubry-André model (AA)~\cite{Aubry80} at the single-particle localization transition. In that case, it was observed that $P_\text{AA}(s) \propto s^{-(1+\Delta_2^\text{AA,gl})}$ with $\Delta_2^\text{AA,gl} \approx0.5$~\cite{Geisel_91} and $K_\text{AA}(\tau) \propto \tau ^{-d_2^\text{AA,gl}}$ with $d_2^\text{AA,gl} \approx0.5$~\cite{hopjan_vidmar_23}, giving $\Delta_2^\text{AA,gl}\approx d_2^\text{AA,gl}$ at the level of the global spectrum.

\clearpage
\end{document}